# Targeting cytochrome C oxidase in mitochondria with Pt(II)-porphyrins for Photodynamic Therapy

Michael Börsch*

3rd Institute of Physics, University of Stuttgart, Pfaffenwaldring 57, 70550 Stuttgart, Germany;

**ABSTRACT**

Mitochondria are the power house of living cells, where the synthesis of the chemical "energy currency" adenosine triphosphate (ATP) occurs. Oxidative phosphorylation by a series of membrane protein complexes I to IV, that is, the electron transport chain, is the source of the electrochemical potential difference or proton motive force (PMF) of protons across the inner mitochondrial membrane. The PMF is required for ATP production by complex V of the electron transport chain, i.e. by $F_oF_1$-ATP synthase. Destroying cytochrome C oxidase (COX; complex IV) in Photodynamic Therapy (PDT) is achieved by the cationic photosensitizer Pt(II)-TMPyP. Electron microscopy revealed the disruption of the mitochondrial christae as a primary step of PDT. Time resolved phosphorescence measurements identified COX as the binding site for Pt(II)-TMPyP in living HeLa cells. As this photosensitizer competed with cytochrome C in binding to COX, destruction of COX might not only disturb ATP synthesis but could expedite the release of cytochrome C to the cytosol inducing apoptosis.

## 1. INTRODUCTION

Targeting the cellular power supply and disrupting the production of the "cellular energy currency" adenosine triphosphate, ATP, is one option to destroy a cancer cell. The mitochondria in mammalian cells are the organelles within the cell, where a series of membrane enzymes controls the synthesis of ATP and the maintenance of appropriate ATP levels. These enzymes are called complex I to V and constitute the electron transport chain for oxidative phosphorylation. Targeting or inhibiting one of these membrane enzymes by specific drugs or destroying them by light-induced local production of highly reactive molecules like Singlet-oxygen, $^1O_2$, will result in fast cell death. To generate $^1O_2$ and related oxygen radicals in cells and tissue by light, chromophores are used as photosensitizers, and this process is called Photodynamic Therapy. Well-known photosensitizers with high quantum yields for $^1O_2$ are the porphyrins and phthalocyanines.

If we want to target the mitochondrial membrane enzymes for Photodynamic Therapy, the photosensitizers have to be transported selectively to the mitochondria after entering the cell. What are the physico-chemical properties of dyes / photosensitizers to ensure transport to the mitochondria ? The group of Herbert W Zimmermann at the University of Freiburg (Germany) has measured transport kinetics of fluorescent dyes to the mitochondria of HeLa cells and found that hydrophobicity and cationic charges of the dyes are the main constituents for transport to the mitochondria [1-7]. Commercially available fluorophores to stain mitochondria selectively like the so-called "Mitotracker"-dyes are also lipophilic (or hydrophobic, respectively) and cationic. The dyes have to be moderately soluble in buffer media to avoid non-specific stacking into all cellular membranes. Accordingly, by modifying porphyrins and phthalocyanines to become cationic and moderately lipophilic, targeting the mitochondria with photosensitizers was achieved [8, 9].

Finally, it has to be unraveled where the photosensitizer bind within the mitochondria. Is it the lipid membrane environment or membrane proteins? Is there a specific target inside mitochondria? Using photo affinity labeling it was shown that only a few membrane proteins were associated with the lipophilic cationic dyes in HeLa cells [6]. The same dye was even more selective in the mitochondria of yeast cells and preferentially attached to cytochrome C oxidase, COX [1].

*m.boersch@physik.uni-stuttgart.de; phone (49) 711 6856 4632; fax (49) 711 6856 5281; www.m-boersch.org

To validate mitochondrial attachment to COX, binding of a cationic and lipophilic pyrene dye was measured *in vivo* using lifetime resolved fluorescence resonance energy transfer, FRET, to the cytochromes in COX as possible acceptor dyes for FRET [2]. Significant shortening of the pyrene lifetime was observed in mitochondria and with reconstituted COX in liposomes *in vitro* indicating the direct binding to complex IV of the electron transport chain.

Here, we report that lipophilic cationic porphyrin-based photosensitizers can also bind directly to COX in mitochondria of living HeLa cells. Time resolved phosphorescence of the Pt(II)-tetramethylpyridylporphyrin Pt(II)-TMPyP (see Fig. 1) is used and shortening of the phosphorescence lifetime upon binding to mitochondria as well as to reconstituted COX in liposomes *in vitro* was detected. Large spectral shifts in absorbance and phosphorescence support a specific binding to a site on COX which competes with binding of cytochrome C. The optical properties of Pt(II)-TMPyP are similar to the more lipophilic porphyrin derivative Pt(II)-TDecPyP with decyl side chains instead of methyl groups.

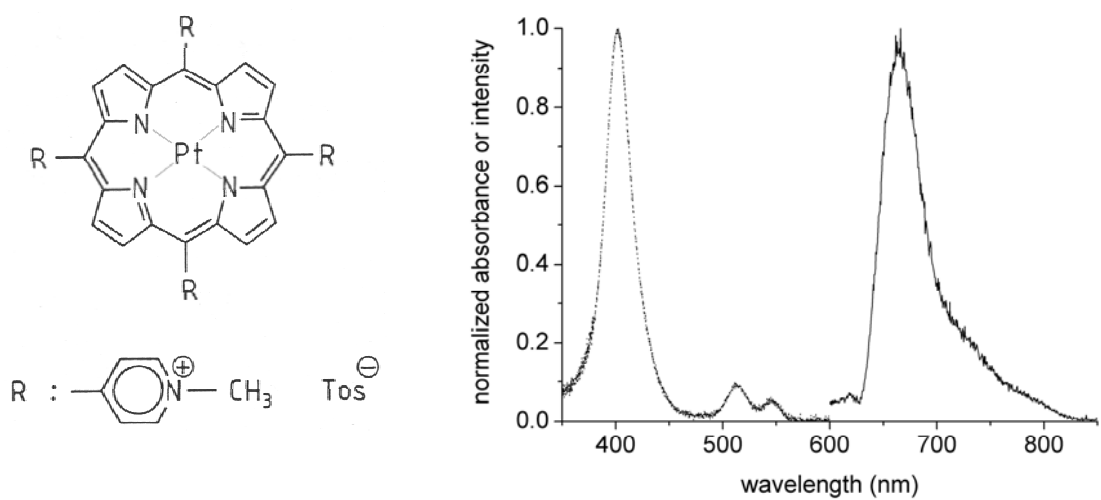

**Fig. 1: A**, **Structure of Pt(II)-TMPyP. B**, UV-VIS absorbance (dotted line, 350 to 600 nm) and phosphorescence spectra (straight line, 600 to 850 nm) of Pt(II)-TMPyP in Earle's balanced salt solution (EBSS) without Phenol Red.

## 2. EXPERIMENTAL PROCEDURES

### 2.1. Synthesis of Pt(II)-TMPyP and Pt(II)-TDecPyP

The synthesis of Pt(II)-TMPyP from Tetrakis(N-methylpyridinium-4-yl)porphyrin ($H_2$TMPyP) was accomplished in aqueous solution according to published procedures [10]. Briefly, Pt(II)$Cl_2$ was dissolved in benzonitrile/DMSO and $H_2O$ in the presence of AgNO$_3$ and stirred for 3 h. AgCl was removed by filtration and $H_2$TMPyP added and heated to reflux for 25 h. After removal of the solvent in vacuum, the red-brownish solid was dissolved in MeOH and filtrated. The orange-red solution contained Pt(II)-TMPyP. Traces of the educt $H_2$TMPyP were removed by successive re-crystallizations in MeOH/ether. Finally Pt(II)-TMPyP was lyophilized and stored at 4°C. Purification of the Pt(II)-porphyrin was checked by thin layer chromatography, UV-VIS absorbance, time resolved phosphorescence and $^1$H-NMR in $D_2O$ at 80°C showing the characteristic pseudo-triplet splitting of the β-pyrrole protons due to $^4$J-coupling to $^{195}$Pt (with a coupling constant 12.3 Hz).

Synthesis of Pt(II)-TDecPyP was achieved in two steps (Fig. 2). First, the Pt(II) complex of Tetrapyridylphorphyrin ($H_2$TPyP) was synthesized similarly to the synthesis of the tetraphenylphorphyrin complex [11]. Pt(II)$Cl_2$ was dissolved in water-free benzonitrile heated to reflux under $N_2$ atmosphere. $H_2$TPyP was added slowly within 21 h using a small Soxleth extraction apparatus. The reaction was monitored by UV-VIS spectroscopy in CHCl$_3$ using the characteristic Soret-band of the educt at 416 nm and the product at 400 nm. The solvent was removed in vacuum. The red-brownish solid was dissolved in CHCl$_3$ and filtrated. Afterwards CHCl$_3$ was removed in vacuum. To purify the Pt(II) complex, the solid was dissolved in 2.5 M HCl and compounds were separated by chromatography (silica-60 column with 0.5 M HCl as eluant). The first dark-green fraction of protonated $H_2$TPyP was discarded. The second orange-red fraction contained Pt(II)-TPyP as a tetrahydrochloride and was immediately shock-frozen with liquid $N_2$ and lyophilized. The tetrahydrochloride of Pt(II)-TPyP was exchanged to the free base by ion exchange chromatography. Therefore, lyophilized tetrahydrochloride of Pt(II)-TPyP was dissolved in freshly distilled pyridine and added to a column with amberlite-410. After elution with pyridine in 10 h, the solvent was removed in vacuum and orange-red microcrystalline Pt(II)-TPyP was obtained. In the second step, Pt(II)-TPyP was reacted with freshly prepared decyl trifoliate in water-free $CH_2Cl_2$ to alkylate the pyridyl groups [12]. Decyl triflate was used in 80-fold excess. The reaction was terminated after for 4 days by removing the solvent in vacuum. Unreacted decyl triflate was separated by chromatography on silanized silica-60-H with $CHCl_3$. The main product was Pt(II)-TDecPyP with all four pyridyl groups alkylated. However, thin-layer chromatography on reversed-phase plates (RP-18 with MeOH containing 5% LiBr as eluant) revealed small amounts of other Pt(II)-TDecPyP derivatives with three, two or only one alkylated pyridyl groups.

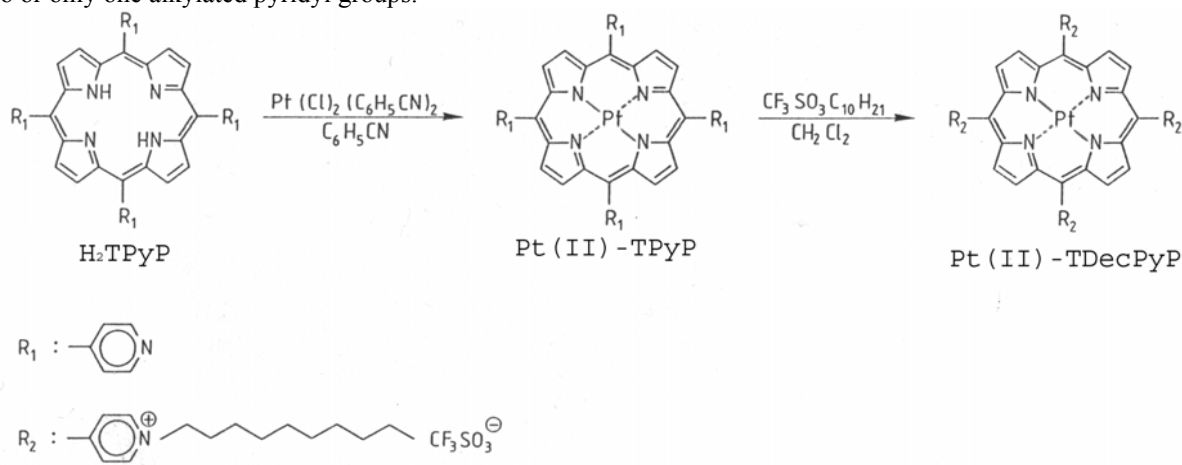

**Fig. 2**. **Synthetic route** for the lipophilic Pt(II)-TDecPyP

## 2.2. Microscope setup for time resolved phosphorescence measurements

Phosphorescence lifetime measurements of the Pt(II)-porphyrins in solution and cells were accomplished by single-shot pulsed excitation using an excimer laser (308 nm, 20 to 45 ns pulses; EMG 201 MSC, Lambda Physik) for pumping a tunable dye laser (FL 2002, Lambda Physik). Laser dyes were PBBO (386 - 420 nm) and Coumarin 120 (423-462 nm). Nanosecond pulses were coupled into a quartz objective (Ultrafluar Zeiss; 100x, N.A. 1.25, glycerol immersion) of the microscope (UEM Zeiss) in epi-fluorescence configuration for confocal or wide field excitation. Alternatively, phosphorescence lifetimes were measured in a home-built quartz cuvette holder with controlled temperature and stirrer. Luminescence was separated from laser light by interference filters (Andovar or Oriel) and detected by a red-sensitive fast photomultiplier (R 1894, Hamamatsu). The analog PMT signals were digitized in 512 channels by a fast digitizer (Tektronics 7912 AD) and stored (Tektronics 4052A). Data conversion to the ASCII format was accomplished *via* an IEEE connection to a box car integrator (4402 EG&G Princeton Applied Research) [2].

Time gated phosphorescence spectra were recorded using the optical multi channel analyzer system (1460 OMA III EG&G Princeton Applied Research) with selectable integration times from 5 ns to 10 ms [13]. Phosphorescence was coupled into a quartz fiber optic and delivered to the spectrograph (MonoSpec 27, Thermo Jarrell-Ash) in front of an intensified, cooled Si detector array (1024 channels). Calibration of the spectrograph was accomplished using spectral lines of a low pressure Hg lamp.

Phosphorescence excitation and emission spectra were measured in quartz cuvettes in a SLM 8100 (SLM Aminco) spectrometer with HITC as a quantum counter to correct for spectral properties of the excitation source.

## 2.3. Staining of HeLa cells, irradiation and electron microscopic imaging of fixed HeLa cells

HeLa cells (Bristol S3) were cultivated in monolayers for 4 days at 37° C in medium consisting of Earle's MEM, 10% foetal calf serum, 1% 0.2 M glutamine, penicillin (100 U $ml^{-1}$) and streptomycin (100 µg $ml^{-1}$) [8]. Cells were washed with Earle's BSS and were incubated with Pt-TMPyP ($1·10^{-6}$ M in Earle's MEM) for 1 h at 37°C. After incubation cells were rinsed with Earle's BSS. To induce PDT cells were irradiated with a 900 W xenon high pressure lamp (Jovy). IR and UV light was filtered by a combination of a 4-cm water absorption cell plus a long-pass filter LP 380 (Oriel). HeLa cells were kept at 37° during the irradiation with 70 mW /$cm^{-2}$. Irradiation power was measured with a calibrated solar cell (Fraunhofer Institut, Freiburg).

For electron microscopy, HeLa cells were grown in a monolayer and were incubated with Pt-TMPyP, washed, irradiated and fixed for 45 min at 5°C in a mixture of osmium tetroxide and 2.5% glutaraldehyde in 0.11 M sodium cacodylate buffer (ph 7.4). After fixation in the dark the cell monolayer was washed three times with distilled water and stained with 25% uranyl acetate solution for 45 min. After dehydration by graded series of EtOH, cells were embedded in HPMA and Epon 812. Polymerized Epon sheets were removed from the culture flask and cut with a diamond knife (Reichert-Jung Ultracut). Slices were put on grids and contrasted with alkaline lead citrate solution for 1 min at 20°C. Cells were imaged on a Zeiss EM 109 electron microscope.

## 3. RESULTS

### 3.1. Localizing Pt(II)-TMPyP in mitochondria of HeLa cells by electron microscopy

Lipophilic cationic dyes accumulate in the mitochondria of living cells and can be detected by fluorescence microscopy. Selective vital staining of mitochondria in HeLa with different types of dyes has been reported, for example with indocarbocyanine derivatives like the photo affinity label azopentylmethylcarbocyanine (APMC) [1,5,6]. This dye binds to four different membrane proteins of the inner mitochondrial membrane. When applied to yeast cells, a single mitochondrial membrane protein was targeted and identified as subunit I of cytochrome C oxidase.

Pt(II)-TMPyP is a tetra-cationic dye with similar hydrophobicity as AMPC. Thus we attempted to localize Pt-TMPyP directly in living HeLa cells by phosphorescence imaging using 400 nm excitation and a photo camera attached to the microscope. However, the low phosphorescence quantum yield of Pt-TMPyP in the presence of $O_2$ prevented a direct visualization. Alternatively, mitochondria can be stained with a bright fluorescent dye in the presence of Pt(II)-porphyrins and monitored using long-wavelength illumination to avoid photosensitizing. Pre-incubating living HeLa cells with Pt-TDecPyP at $10^{-6}$ M for 15 min and subsequent staining with 1,1'-Dihexyl-3,3,3',3'-tetramethylindocarbocyanine IC6 ($5·10^{-8}$ M for 5 min; Molecular Probes) unraveled a shorting of the mitochondrial structures, which indicates cytotoxic effects. When Pt-TDecPyP-incubated HeLa cells were irradiated with 70 mW $cm^{-2}$ for 15 min and subsequently stained with IC6, the shape of the mitochondria appeared to be strongly altered and globular. However, double-staining provides only indirect evidence for a mitochondrial accumulation of the Pt(II)-porphyrins.

Electron microscopic (EM) images provide a detailed view on the cellular compartments and membranes with about 2 nm resolution. To identify the accumulation location of Pt-TMPyP as a photosensitizer we compared EM images of untreated HeLa cells prior to and after irradiation with 70 mW $cm^{-2}$ with Pt-TMPyP-treated HeLa cells (Fig. 3). In the absence of the photosensitizer, irradiation of the cells did not alter the filamentous ultra structure of the mitochondrial christae (Fig. 3 A and B). The endoplasmatic reticulum (ER), ribosomes as well as the nuclear membranes plasma membrane remained unchanged upon irradiation. Vital staining with $10^{-6}$ M Pt-TMPyP for 15 h did not affect the shape of the mitochondria nor affected the ER or the nucleus (Fig. 3 C). However, after irradiation in the presence of the Pt-porphyrin, cells were massively damaged (Fig. 3 D). Mitochondria appear as globular swollen particles with internal electron-dense granula, and the christae are lost. The ER is strongly enlarged and swollen, and nuclear membrane is partly detached from the nucleus. The nucleus contains electron-dense parts. The morphological changes are very similar to those observed with $H_2$TMPyP. From EM images of HeLa cells irradiated for only 5 min (images not shown), it could be concluded that the photodynamic action starts from the mitochondria. *Vice versa,* Pt(II)-TMPyP is targeting the mitochondria.

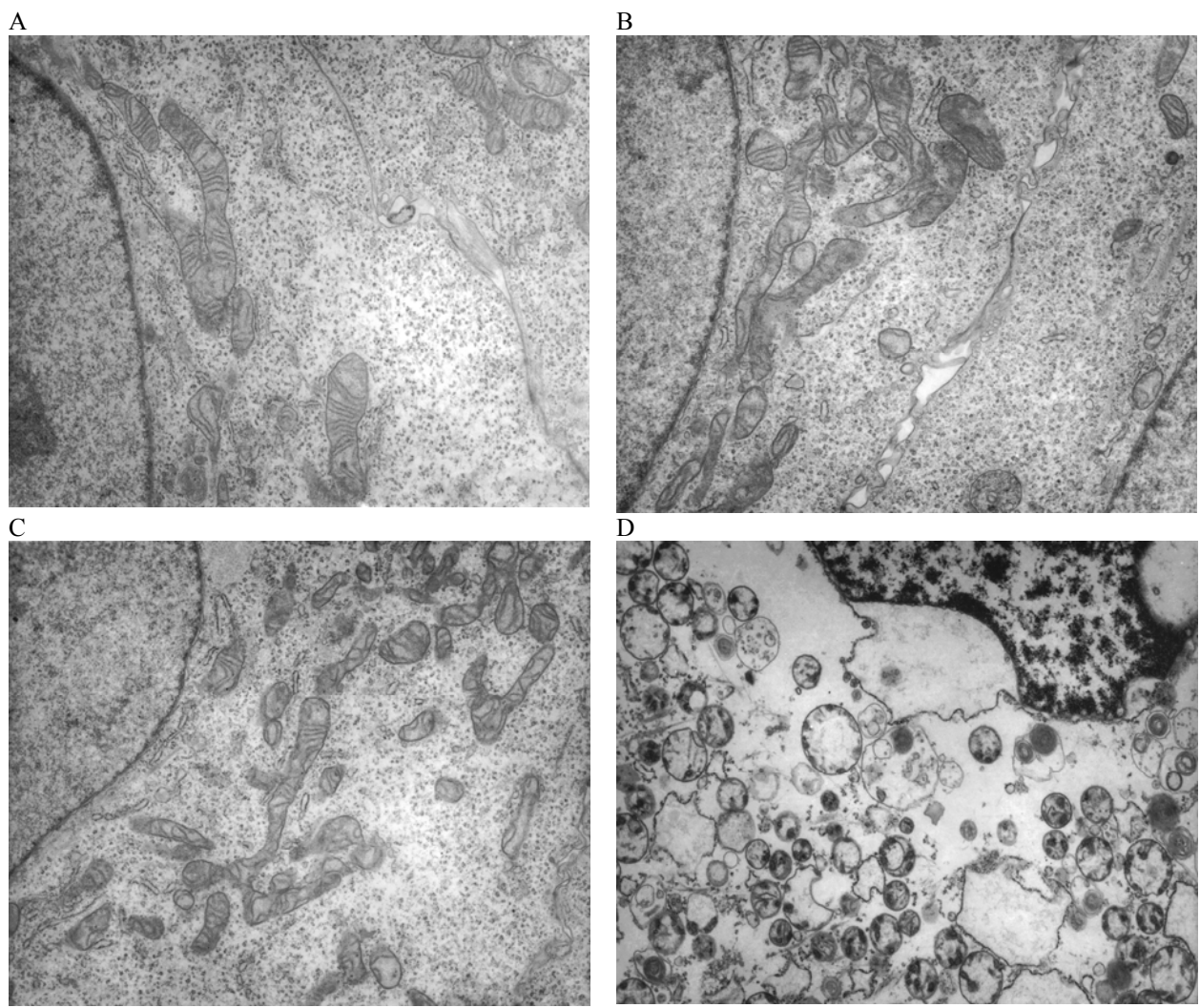

**Fig. 3: Electron micrographs of HeLa cells** after negative staining (7000-fold magnification). **A, B**: HeLa in the absence of Pt(II)-TMPyP; **C, D**: HeLa incubated with $10^{-6}$ M Pt-TMPyP (lower row). **A, C**: before irradiation; **B, D** (right column): after 15 min irradiation with 70 mW/cm$^2$ using a long-pass filter LP 380.

The photodynamic activity of Pt(II)-TMPyP was checked independently by $O_2$ consumption measurements of suspended HeLa cells. Alternatively, counting dead cells by Trypan Blue staining after irradiation revealed a slightly reduced PDT activity compared to $H_2$TMPyP or $H_2$TDecPyP ("POR10" [8] ).

### 3.2. Spectroscopy of Pt(II)-TMPyP in living HeLa cells

The UV-VIS absorbance spectra of Pt(II)-tetraphenylporphyrin (Pt-TPP), Pt(II)-tetrapyridylporphyrin Pt-TPyP and Pt(II)-TMPyP are characterized by three bands. The $S_0 \rightarrow S_1$ transition is spilt into two Q-bands around 510 nm ($Q_{1,0}$) and 540 nm ($Q_{0,0}$) with low oscillator strength. The $S_0 \rightarrow S_2$ transition (Soret band, or B-band) with high oscillator strength is observed around 400 nm. Phosphorescence of these Pt-porphyrins is observed with a maximum around 670 nm ($T_{0,0}$) and a vibronic sideband ($T_{0,1}$) around 730 nm (see Fig. 4, Pt(II)-TMPyP in buffer solution EBSS).

After incubating living HeLa cells with 1 μM Pt(II)-TMPyP for 15 min and subsequent washing steps, direct localization of the porphyrins by phosphorescence imaging in a conventional fluorescence camera microscope was not successful. However, Pt-porphyrins could be detected by cuvette-based spectroscopy. Phosphorescence excitation and emission spectra of Pt(II)-TMPyP-stained living HeLa cell suspensions showed large spectral shifts compared to solution in EBSS buffer. The Soret band was found at 419 nm (i.e. +17 nm shifted to longer wavelength, "red shifted"). $Q_{(1,0)}$ was shifted to 523 nm (+9 nm) and $Q_{(0,0)}$ to 554 nm (+9 nm). The relative oscillator strength ratio Soret/$Q_{(1,0)}$ diminished to 3.4 in contrast to 9.2 in EBSS buffer (Fig. 4, straight black and dotted curves). The phosphorescence maximum in living HeLa was red-shifted by 8 nm to 673 nm. What is the subcellular environment for Pt(II)-TMPyP in HeLa cells that causes these strong interactions resulting in bathochromic shifts and hypochromic effects on the ratio Soret/$Q_{(1,0)}$?

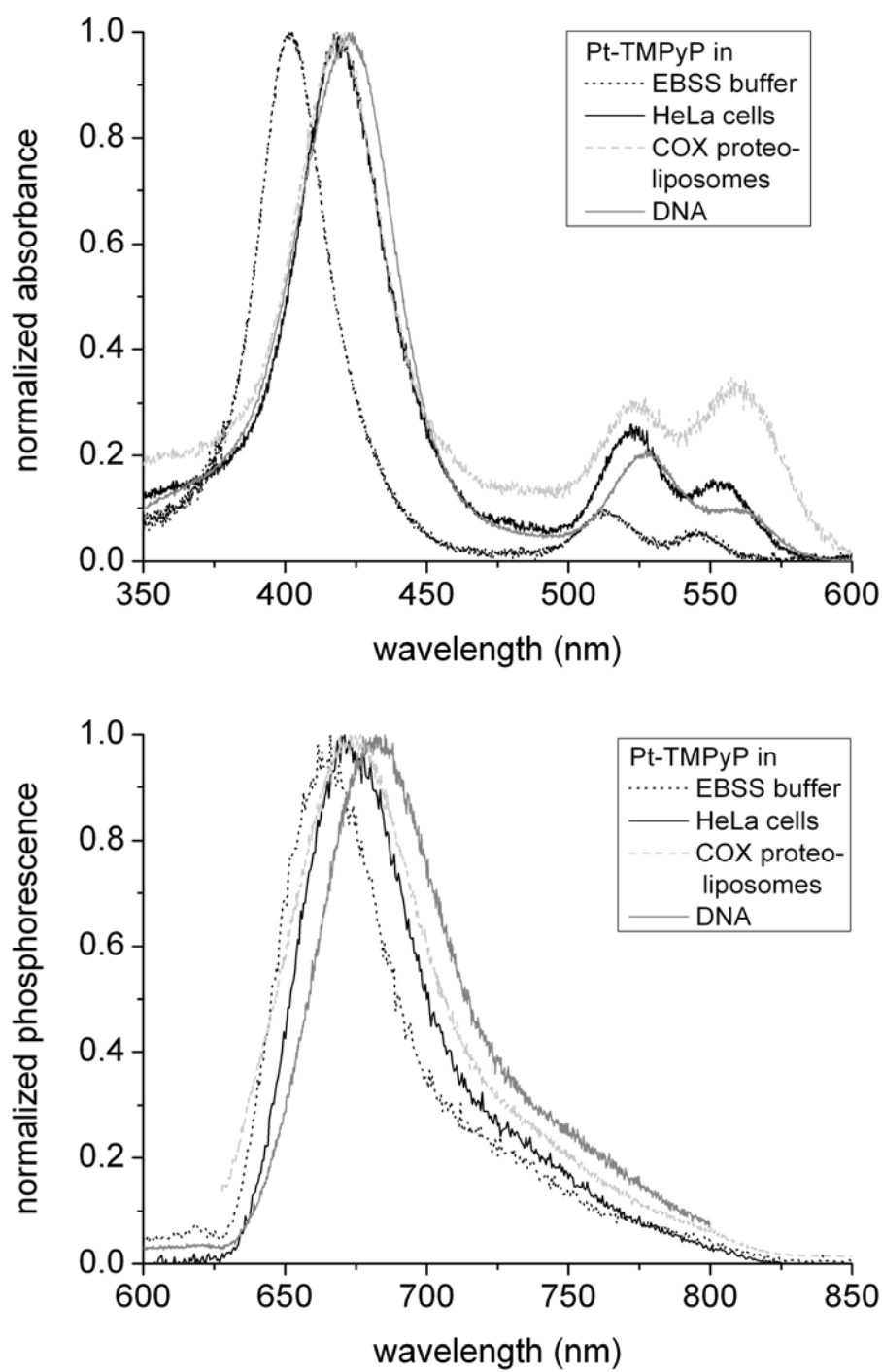

**Fig. 4: Phosphorescence excitation and emission spectra of Pt(II)-TMPyP in living HeLa cells**, in EBSS buffer, bound to DNA and to COX proteoliposomes. Upper panel: excitation spectra. Lower panel: emission spectra.

To identify the binding sites of Pt(II)-TMPyP in living HeLa or the mitochondria, respectively, several distinct scenarios were considered. At first, binding to DNA in fixed HeLa cells was investigated. Following EtOH treatment of HeLa cells, the cationic Pt-porphyrins bound to DNA in the nucleus and direct phosphorescence imaging was possible (to be published elsewhere). The excitation and emission spectra of Pt(II)-TMPyP bound to purified λ-DNA *in vitro* showed large red shifts (Soret band at 423 nm) indicating strong binding (Fig. 4). The phosphorescence maximum was shifted to 683 nm. However, the Soret/$Q_{(1,0)}$ ratio of 6.5 showed a significantly weaker hypochromic effect. Good agreement with the spectral shifts of Pt(II)-TMPyP in living HeLa was found after binding to mitochondria suspensions, whereas binding to lipid vesicles did not alter the spectral properties of Pt(II)-TMPyP (data not shown). The best mimicking environment was proteoliposomes with reconstituted cytochrome C oxidase COX. After mixing COX liposomes with Pt(II)-TMPyP and dialysis for 20 h, the bound Pt-porphyrins showed the same maxima in excitation and emission spectra as in living HeLa, and the Soret/$Q_{(1,0)}$ ratio of 3.5 showed the same strong hypochromic effect (Fig. 4).

### 3.3. Time resolved spectroscopy of Pt(II)-TMPyP in HeLa cells

Altered spectra of Pt(II)-TMPyP bound in living HeLa cells were accompanied by phosphorescence lifetime changes. Lifetimes were recorded using single-shot pulsed excitation with a excimer laser-pumped dye laser tuned to the maximum of the Soret band, and by fast analog-to-digital conversion of the photomultiplier signal. The excited triplet state of Pt-porphyrins is known to be quenched by $O_2$. The phosphorescence lifetime of Pt(II)-TMPyP in $O_2$-saturated EtOH was $\tau = 260$ ns and in $H_2O$ $\tau = 960$ ns (Fig. 5, first row) according to monoexponential decay fitting. In degassed $O_2$-free EtOH, the lifetime of Pt(II)-TMPyP was $\tau = 42$ μs comparable to the phosphorescence lifetime of Pt(II)-TPP in $O_2$-free THF ($\tau = 53$ μs; in good agreement with the literature [14]). When bound in living HeLa, the phosphorescence lifetime decay was multi exponential (Fig. 5, second row). Lifetime components of 1.2 μs, 5.8 μs and longer than 25 μs were interpreted as buffer environment and binding to proteins or DNA which prevented $O_2$ quenching. The dominant lifetime component was $\tau = 190$ ns and therefore significantly shorter than in oxygenated buffer. This lifetime was found following 418 nm excitation, but was not detected after 402 nm excitation.

The DNA environment in fixed HeLa cells resulted in a long phosphorescence lifetime of $\tau = 8.1$ μs and in λ-DNA solution $\tau = 8\pm1$ ns. Phosphorescence lifetime could be fitted monoexponentially. Importantly, a short lifetime component as in the case if living HeLa was not detectable (Fig. 5, third and forth rows). In contrast, after binding to reconstituted cytochrome C oxidase the phosphorescence lifetime of Pt(II)-TMPyP was $\tau = 210\pm20$ ns (with an amplitude of 75%) and a second component $\tau = 1.3\pm0.2$ μs (25%). After irradiation with 70 mW/cm$^2$ to start photodynamic activity the amplitudes of the phosphorescence lifetimes were changed. The amplitude of the long lifetime component around 1 μs increased to 60% indicating the release from COX into the buffer (Fig. 5, lower row).

### 4. DISCUSSION

Cationic Pt(II)-porphyrins like Pt(II)-TMPyP and Pt(II)-TDecPyP are taken up by living cancer cells. They act as photosensitizers to produce $^1O_2$. Electron microscopic images revealed the damaging of mitochondria in HeLa cells after irradiation with 70 mW/cm$^2$ with morphology changes similar to the photodynamic action of the related metal-free porphyrin derivatives [8]. Pt(II)-porphyrins are known to exhibit relatively strong phosphorescence and neglectable E-type delayed fluorescence[15, 16] making them suitable for time resolved spectroscopy and imaging. Phosphorescence lifetime is quenched by molecular oxygen $^3O_2$ and, accordingly, Pt(II)-porphyrins are used as oxygen sensors in polymeric form[17] or bound to protein layers[18].

In this study, the properties of Pt(II)-TMPyP as a photosensitizer and as a phosphorescence marker for a specific environment in the mitochondria are evaluated. The aim was to simultaneously induce and monitor and localize the sites of Photodynamic Therapy in HeLa cells. However, the sensitivity of a photo camera as the detector on the microscope was insufficient to localize the Pt-porphyrin in the mitochondria. Advanced EMCCD cameras with single photon detection capabilities might be sensitive enough to observe PDT by Pt(II)-TMPyP in real time.

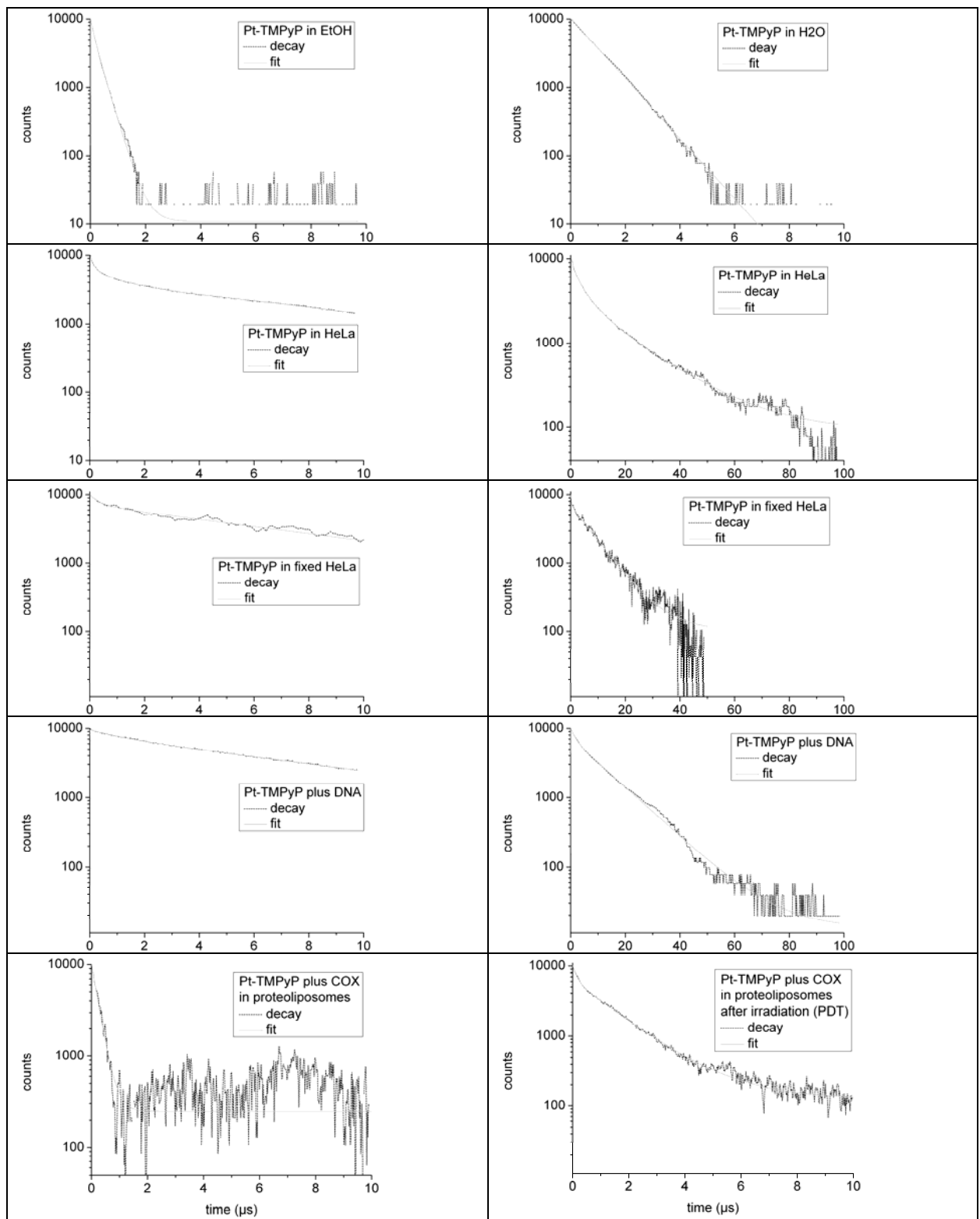

**Fig. 5: Phosphorescence decay curves of Pt(II)-TMPyP** in different environments. Decays in EtOH and H$_2$O in the first row, in living HeLa cells in the second row, in fixed HeLa cells in the third row, in the presence of DNA in the forth row, and bound to reconstituted cytochrome C oxidase in the lowest, fifth row. Notice the different time windows (10 µs or 100 µs). Fitting curves are shown as grey straight lines.

For prolonged observation of PDT in cells, reaching a single-molecule sensitivity with phosphorescent photosensitizers would allow to reduce cytotoxic effects and to discriminate cytotoxicity from photodynamic destruction effects. For confocal single-molecule imaging and localization in living cells, bright fluorophores with exceptional photo stability are required[19]. Thus new ultrastable dyes and fluorescent nanocrystals are continuously developed[20-23]. These fluorophores emit bursts of photons, which yield diffusion properties, association status and conformational changes *in vitro* and *in vivo* by fluorescence correlation spectroscopy[24-28]. To generate burst of photons during transit through the laser focus, the fluorescence lifetime has to be in the nanosecond time range. If we consider the long phosphorescence lifetimes in the micro- to millisecond time range of the Pt(II)-porphyrins, single Pt(II)-TMPyP detection seems to be impossible.

However, when bound to reconstituted cytochrome C oxidase *in vitro*, the phosphorescence lifetime of Pt(II)-TMPyP was shortened. This lifetime information was essential to discriminate the binding site on the cytochrome-containing protein from a DNA binding. DNA binding in vitro and in fixed cells caused similar strong spectral shifts in the static spectra, but did not exhibit the short lifetime component. As other biological environments like proteins or lipids did not result in specific changes of the optical properties of Pt(II)-TMPyP, the question raised about photophysical nature of the strong interaction. Two possibilities were considered: resonance energy transfer (RET) to non-fluorescent cytochromes in COX as RET acceptors or reversible electron transfer / triplet state quenching from COX. Fluorescence resonance energy transfer (FRET) to unravel association of proteins[29-31] or specific binding of ligands[32], or to monitor conformational changes like in the rotary biological nanomotor $F_oF_1$-ATP synthase on the single-molecule level[33-35] are most successful with an fluorescent acceptor for ratiometric data analysis. If the acceptor is a quencher only, donor lifetime changes can be used to indicate the existence of FRET.

A possible photo induced electron transfer causing the shortening of the Pt(II)-TMPyP phosphorescence lifetime bound to COX is supported by binding experiments of tetra-anionic phthalocyanines and porphyrins to the redox counterpart cytochrome C. The tetra-anionic porphyrin TCPP binds specifically to cytochrome C [36] in an 1:1 complex and causes strong hypochromic effects. Four lysines in a rectangular distribution on the surface of cytochrome C provide the electrostatic pattern for oriented binding of the tetra-anionic porphyrin resulting in binding constants lower than 5 μM. The charge distribution of TCPP mimics the negative charges on the cytochrome C binding site on COX. *Vice versa*, it can be anticipated that the tetra-cationic Pt(II)-TMPyP could similarly bind to COX at the cytochrome C binding site. When bound to cytochrome C, the tetra-anionic aluminium phthalocyanine tetrasulfonate showed reversible triplet state quenching [37]. Here, photoinduced electron transfer from the cytochrome quenched the triplet state of the phthalocyanine with electron-transfer rate constants between $10^7$ s$^{-1}$ and $10^8$ s$^{-1}$, i.e. relaxation times of the triplet state in the range of 100 nanoseconds. If Pt(II)-TMPyP binds similarly to COX and electron-transfer rate constants are in a similar order of magnitude, photoinduced electron transfer could explain the specific shortening of the phosphorescence lifetime or triplet state of the Pt(II)-porphyrin.

**Acknowledgements**
The author want to thank H. W. Zimmermann for enabling this work at the Institute for Physical Chemistry, University of Freiburg, Germany. I am grateful to T. Cernay, H. Dummin, H. Haass-Männle and J. R. Grezes for their help in developing the Pt(II)-porphyrin synthesis and for comparison to other photosensitizers located in the mitochondria. A. Naujok, K.-O. Lorenz and T. Berthold accomplished the cell biology and operated the electron microscopy in excellent ways. I thank H. Hüglin, W. Pachaly and W. Seifert for providing the setup of a confocal microscope with pulsed excimer and dye lasers as well as the time resolved detection scheme based on single-shot analog-to-digital PMT signal conversion, P. Gedeck for lifetime fitting software, and F. Männle for NMR spectroscopy.